\documentclass{article}

\usepackage{PRIMEarxiv}

\usepackage[utf8]{inputenc} 
\usepackage[T1]{fontenc}    
\usepackage{hyperref}       
\usepackage{url}            
\usepackage{booktabs}       
\usepackage{amsfonts}       
\usepackage{nicefrac}       
\usepackage{microtype}      
\usepackage{lipsum}
\usepackage{fancyhdr}       
\usepackage{graphicx}       
\usepackage{subfiles}
\usepackage{seqsplit}
\usepackage{amsmath, amssymb, amsfonts}
\usepackage[adjustbox]{adjustbox}

\graphicspath{{figures/}}     

\title{PriGen: Towards Automated Translation of Android Applications' Code to Privacy Captions
}

\author{
  Vijayanta Jain, \\
  University of Maine \\
  Orono, ME\\
  \texttt{vijayanta.jain@maine.edu} \\
   \And
   Sanonda Datta Gupta, \\
   University of Maine \\
   Orono, ME\\
   \texttt{sanonda.gupta@maine.edu} \\
  \And
  Sepideh Ghanavati\\
  University of Maine \\
  Orono, ME\\
  \texttt{sepideh.ghanavati@maine.edu} \\
  \And
  Sai Teja Peddinti \\
  Google Inc. \\
  Mountain View, CA\\
  \texttt{psaiteja@google.com} \\}

\begin{document}
\maketitle

\begin{abstract}
Mobile applications are required to give privacy notices to the users when they collect or share personal information. Creating consistent and concise privacy notices can be a challenging task for developers. Previous work has attempted to help developers create privacy notices through a questionnaire or predefined templates. In this paper, we propose a novel approach and a framework, called PriGen, that extends these prior work. PriGen uses static analysis to identify Android applications' code segments which process sensitive information (i.e. \textit{permission-requiring code segments}) and then leverages a Neural Machine Translation model to translate them into \textit{privacy captions}. We present the initial evaluation of our translation task for $\sim$300,000 code segments.

\end{abstract}


\section{Introduction}

Privacy notices are defined as an application's artifact that inform the users about the collection and processing of their personal information. Since mobile applications may use sensitive personal information for their core functionalities as well as for advertisements and performance assessments~\cite{peddinti2019reducing}, the Federal Trade Commission recommends companies to inform users about their data practices and give them choices~\cite{federal2013mobile}. Currently, three distinguished mechanisms exist for providing privacy notices for mobile applications: (i) a privacy policy, (ii) an application description, and (iii) a permission rationale. Although these mechanisms seem adequate, the generated notices might be inconsistent, generic, or incomplete~\cite{liu2018large,okoyomon2019ridiculousness,sun2020quality,zimmeck2019maps} and they might remain static when the code changes.

In recent years, much effort has been done to identify inconsistencies between mobile applications and their privacy notices~\cite{gorla2014checking,liu2018large,okoyomon2019ridiculousness,reyes2018won,slavin2016pvdetector,zimmeck2019maps}. Other research attempts to create privacy policies, notices, or permission-based descriptions by using a questionnaire \cite{liu2018mining,rosen2013appprofiler,rowan2014encouraging}, and/or by evaluating the code behavior~\cite{yu2016toward,zimmeckEtAlPrivacyFlashPro2021}. While the above work are promising and have shown initial success, they either rely on applications' descriptions which might not fully describe the code's behavior and lead to inconsistencies
\cite{liu2018mining,rowan2014encouraging} or they rely on predefined templates \cite{yu2016toward} which may result in generic notices without including any specific rationale for using sensitive information. More importantly, these approaches do not provide traceability between the source code and the generated privacy statements which can result in inconsistencies when the source code changes.

To address these challenges and to help developers provide privacy notices with rationales that match applications' source code, even when a change occurs, we propose a framework called \textit{PriGen} (pronounced \textit{pry-gen}). PriGen provides deep learning models, methodologies, and tools to help developers generate privacy captions directly from an application's source code. We define \textit{privacy captions} as concise sentences that accurately describe an application's privacy practices. The generated privacy captions can serve four distinct purposes: First, they can be revised and included in different privacy notice formats; second, they can be used as privacy engineering guidelines for developers to help them improve their code; third, they can be used in discussions between developers, legal and policy experts, and business executives while creating privacy notices; and fourth, they can be used for internal and external audits as evidence of application's privacy-preserving practices. The purposes of privacy caption provide an overview of how PriGen presents additional advantages over other related work (a detailed comparison with related work is described in Section~\ref{sec:relatedwork}).

We posit that generating privacy captions from code is a language translation problem, where the source language is an extracted \textit{permission-requiring} code segment (PRCS) (i.e. a type of method which processes sensitive information), and the target language is a privacy caption that describes the code segment. A PRCS can include multiple methods that are nodes of a call graph. This is because the first method in the call graph (first hop) that processes a sensitive information can share that information with other methods (second, third, or further hops) which also process them. In this paper, we only consider the first hop (a single method) that calls a \textit{permission-requiring} API. PriGen extends the current effort in generating comments and commit messages from source code \cite{alon2018code2seq,leclair2020improved,loyola2017neural,jiang2017automatically} with Neural Machine Translation (NMT) models \cite{bahdanau2014neural} to privacy engineering domain for generating privacy captions.


\begin{figure*}[htp!]
    \centering
    \fbox{\includegraphics[width=0.9\textwidth]{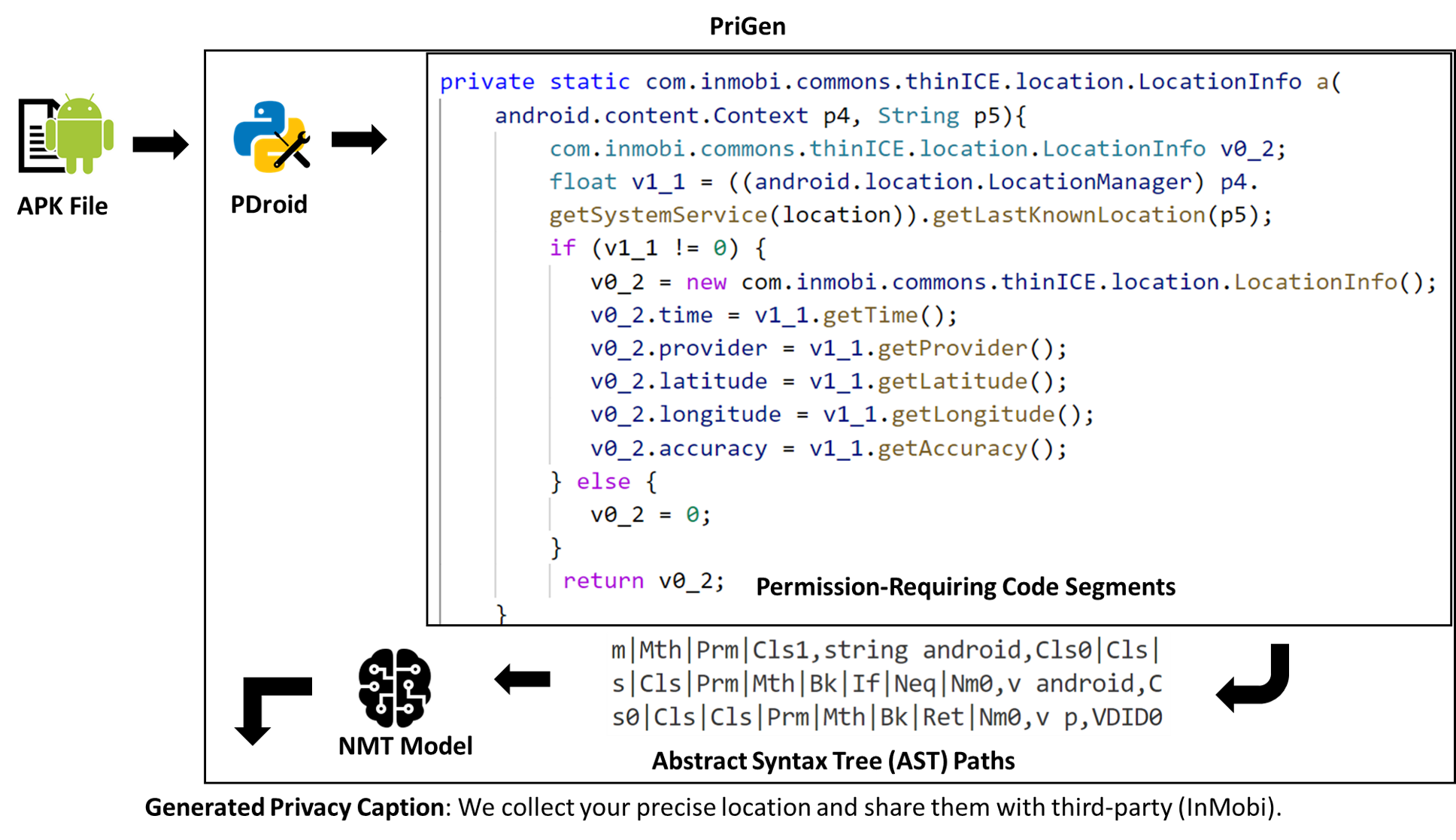}}
    \caption{A High-Level Overview of the PriGen Framework}
    \label{fig:overview}
\end{figure*}

In this paper, we describe PriGen's functionalities, our novel approach to generate privacy captions, and our static analysis approach with its tool, \textit{PDroid}\footnote{https://www.github.com/vijayantajain/PDroid} (Section~\ref{sec:methodology}). We also present our preliminary results for generating privacy captions and discuss challenges and future work (Section~\ref{sec:discussion}). To show initial results, we downloaded $\sim$80,000 Android applications from the Androzoo collection~\cite{Allix:2016:ACM:2901739.2903508}, created a dataset of $\sim$300,000 PRCS, and finally trained Code2Seq\footnote{https://github.com/tech-srl/code2seq} ~\cite{alon2018code2seq}, a state-of-the-art NMT model, to generate captions. We leverage Code2Seqfor our PRCS translation task as the code is open-sourced and easy to experiment with other datasets. Our preliminary results show that, using an NMT model, we can generate accurate and readable captions for a PRCS with 4-5 lines of code.

\section{Related Work}

\subsubsection{Privacy Consistency Analysis:} 
Much research focuses on identifying inconsistencies between an application's source code and its privacy  notices~\cite{okoyomon2019ridiculousness,zimmeck2019maps,liu2018large,gorla2014checking}. Zimmeck et al. \cite{zimmeck2019maps} evaluate over a million Android applications and show an average of 2.89 inconsistencies between an application and its privacy policy. Gorla et al. \cite{gorla2014checking} and Liu et al. \cite{liu2018large} identify mismatches between Android applications' behavior and their descriptions or permission rationales. While these work are promising, they do not provide any support to resolve the inconsistencies. Our work differs from these approaches in that PriGen helps \textit{resolve} inconsistencies by offering traceability between code and privacy captions (and therefore privacy notices) to determine where changes need to happen (either in code or in the notices).

\subsubsection{Generating Privacy Notices:} 

Other approaches attempt to improve the understanding of applications' privacy practices by generating permission explaining sentences~\cite{liu2018mining} or template-based privacy policies~\cite{rowan2014encouraging}. PAGE \cite{rowan2014encouraging} and AppProfiler \cite{rosen2013appprofiler} generate privacy policies from a set of questions without considering the source code. CLAP mines similar applications' descriptions, identifies reasons behind the required permissions, and summarizes them to generate permission explaining sentences \cite{liu2018mining}.  AutoPPG \cite{yu2016toward}, a closely related work, creates a privacy policy by analyzing the code and using the predefined \texttt{subject form object [condition]} format to create privacy statements. This approach provides generic privacy statements that do not contain any rationale. PriGen differs from AutoPPG \cite{yu2016toward} in that it generates captions that provide rationales for using sensitive information. This is because the NMT model of PriGen is trained on a dataset of PRCS, which contains multiple hops, and includes human-annotated privacy captions with rationales. Figure \ref{fig:overview} shows a PRCS with a privacy caption that the NMT model generates. PriGen generates multiple captions for an Android application which developers can combine to create a single privacy notice such as \emph{``We collect location to find nearby cabs and to share it with third-parties"}. This is a more comprehensive caption as compared to what AutoPPG can generate  \emph{``We would use your location (including, latitude and longitude)"} \cite{yu2016toward}. PrivacyFlashPro \cite{zimmeckEtAlPrivacyFlashPro2021}, another closely related work, is a tool that combines code analysis with a questionnaire to generate legally compliant privacy policies for iOS applications. PriGen differs from PrivacyFlashPro since it provides traceability, i.e., it shows which code segments are translated to which privacy captions. This kind of traceability not only supports consistency between an application's source code and its privacy notice but also enhances internal and external legal or policy audits.

\subsubsection{Translating Code to Natural Language Sentences:} 

NMT models have been widely used in software engineering research to generate comments for the source code~\cite{alon2018code2seq,iyer2016summarizing,leclair2020improved} or commit messages for the changes in a code \cite{loyola2017neural,jiang2017automatically}. These works use an encoder-decoder architecture where an encoder creates an internal representation of the code segment and the decoder translates this representation into natural language statement. Alon et al. \cite{alon2018code2seq} propose Code2Seq, an NMT model based on the Seq2Seq architecture to generate code captions. This approach provides one of the best scores for the BLEU metric in captioning code as compared to other similar models. LeClair et al. \cite{leclair2020improved} propose to use Convolutional Graph Neural Networks(ConvGNN) to translate code into natural language and achieves better BLEU-4 and ROUGE-LCS scores than Code2Seq.

\label{sec:relatedwork} 

\section{The PriGen Framework}

\subsection{Overview of PriGen}
\label{subsec:prigen_overview}

PriGen framework includes two functionalities: (1) Extracting the PRCS data from an Android application and (2) Generating privacy captions with their rationale. For this, it first takes an Android Application Package (APK) file, and by using PDroid, our static analysis tool, it identifies and extracts the \textit{Permission-Requiring} Code Segments (PRCS). Next, it preprocesses the PRCS by generating their Abstract Syntax Tree (AST) paths. Finally, it uses a trained NMT model to predict a privacy caption for each code segment from its AST paths. Figure~\ref{fig:overview} shows a high-level overview of PriGen. 

\subsection{Creating the Permission-Requiring Code Segments Dataset}
\label{subsec:creating_dataset}
We create a PRCS dataset from $\sim$80,000 APK files which we downloaded from the Androzoo Collection~\cite{Allix:2016:ACM:2901739.2903508}. For this, we (a) identify and extract the PRCS; and (b) create privacy descriptions for the PRCS.

\subsubsection{Identifying and Extracting the PRCS:}
To protect the users' privacy, access to users' sensitive information is governed by system APIs and permissions in Android. If an Android application wants to access users' sensitive information, it must call a system API and declare the necessary permissions in \texttt{AndroidManifest.xml} file. We call the system APIs which require permissions, \textit{permission-requiring} APIs. We leverage a static analysis approach to first identify the \textit{permission-requiring} APIs and then extract the methods that call these APIs (i.e. first hop in a PRCS).  In future, we will extract multiple hops. 

To identify \textit{permission-requiring} APIs, we manually search the Android Developer Documentation\footnote{https://developer.android.com/reference/} and extract the API name, its class name, its description string, and the sensitive information accessed. We save this information in a JSON file. We also consider deprecated APIs from the documentation, since some Android applications might be built for older Android SDKs. In this work, we only include 69 APIs that require permissions related to Internet, Network, and Location permission groups. We will develop a thorough mapping of all APIs and permissions, in future. 

We created PDroid, a static analysis tool, using Androguard~\cite{androguard}, that analyzes an APK to identify and extract the PRCS. PDroid uses the JSON file created above to locate all the \textit{permission-requiring} APIs that are used in an APK file. Then, using Androguard's call graph analysis, it identifies all the methods that call at least one \textit{permission-requiring} API and extracts their source code. Using PDroid, we extracted $\sim$300,000 PRCS from the $\sim$80K downloaded APK files.

\subsubsection{Creating Privacy Captions for the PRCS:}
After extracting the PRCS, we create a privacy caption for each PRCS. Since manually creating such sentences for a large dataset is not feasible, we propose a semi-automated approach using Human-in-the-Loop framework. For this, we first generate \textit{code} captions of a PRCS using a pre-trained Code2Seq~\cite{alon2018code2seq} model, and then concatenate these captions with the API descriptions for those \textit{permission-requiring} APIs that are called in the PRCS to create privacy captions. A code caption tells us \textit{``what does this code do?''}, whereas the API description describes \textit{``the nature of the sensitive information accessed''}. Combining the two should approximately help us answer \textit{``how does this code process the sensitive information?''}. After concatenation, we evaluate the statements for accuracy and coherence and then, manually modify them to improve their quality. In this work, we only focus on the code captioning step and the Human-in-the-Loop step is left for future work.

To this end, we have trained Code2Seq~\cite{alon2018code2seq} on Funcom dataset~\cite{leclair2019recommendations} and generated code captions for the PRCS. We choose Funcom as our code captioning dataset, instead of Code2Seq's own Java Large dataset~\cite{alon2018code2seq}, because the captions in Funcom dataset are much longer than in Java Large and the dataset is still considerably large with $\sim$2 million examples. We train Code2Seq model with the following configuration parameters\footnote{https://github.com/tech-srl/code2seq/blob/master/config.py}: Embedding Size - 512, RNN Size - 512, and Decoder Size - 512, and Max Target Parts - 37.

\label{sec:methodology} 

\section{Results and Discussion}

We quantitatively evaluate our model (i.e. Code2Seq trained on Funcom) with cumulative BLEU-4 and ROUGE LCS scores. These scores are calculated using a validation set containing 10\% samples of Funcom. We achieved a cumulative BLEU-4 score of 18.9 and ROUGE LCS scores for precision, recall, and F1 of 47.71, 44.46, and 44.49 respectively. Comparing our results with ConvGNN \cite{leclair2020improved}, we find that our model performed reasonably well. ConvGNN approach achieved a BLEU score of 19.93 which is close to ours. Our ROUGE-F1 score is lower by almost 10 points because of the difference in the two models' architecture. ConvGNN architecture additionally leverages the graph structure of AST paths and this improves the scores for the code captioning task.

\begin{figure*}
    \centering
    \fbox{\includegraphics[width=1\textwidth,]{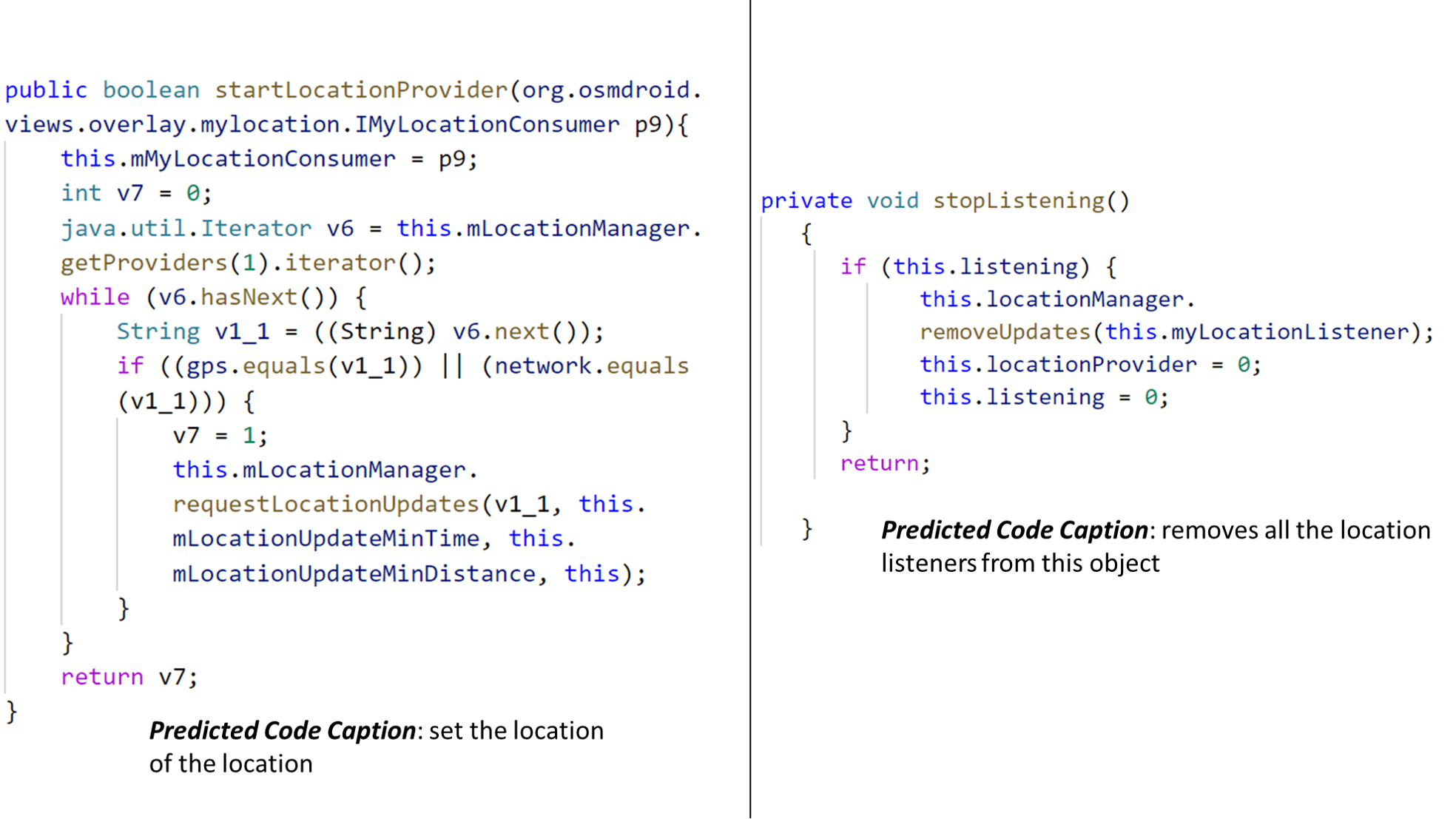}}
    \caption{An Example of Two PRCS with Their Code Captions Predicted by the Trained Code2Seq Model.}
    \label{fig:qualitative_evaluation}
\end{figure*}

In our preliminary qualitative analysis of Code2Seq model, we randomly selected 50 code samples from the PRCS dataset and used the trained Code2Seq model to generate their code captions. The first two authors then manually analyzed the predicted code captions for accuracy and readability. For accuracy, we read each caption and then evaluated if it can accurately describe the code segment (a binary response). For readability, we looked for grammatical correctness and lack of repetition of words and phrases. We found that the predicted code captions were accurate and readable for code segments with 4-5 lines of code (LOC), which is a promising result. However, for longer code segments (i.e. LOC $> 5$), the captions were vague and often contained repetitive words. As shown in Figure~\ref{fig:qualitative_evaluation}, for the PRCS on the right, our model predicts an accurate and readable code caption, while the predicted code caption for the PRCS on the left is vague and includes repeated words. We plan to do a formal qualitative analysis with larger number of samples at a later stage.

\subsection{Current Challenges}
\label{subsec:challenges}

We plan to resolve the following challenges in future:

1. The captions predicted by our model are accurate and readable for code segments with 4-5 LOC. To be able to translate longer code segments, our plan is to explore other NMT architectures, such as ConvGNN. Currently, we are also manually modifying the generated captions for about 30,000 PRCS which will be used to re-train our NMT model.

2. Code reuse is a common practice in software engineering and developers often use code segments obtained from the internet to develop applications. Because of this practice, there are several PRCS that are similar in our dataset which can cause an NMT model to overfit. In future, we will use software similarity to identify and remove duplicate samples.

3. The PRCS dataset contains several obfuscated methods which means that the original identifier names are replaced with generic ones. The NMT model cannot generate meaningful code captions for obfuscated code. To address this challenge, we will remove obfuscated code segments from our dataset, since developers, will use the PriGen's model on their un-obfuscated code.

\label{sec:discussion} 

\section{Conclusion}

In this paper, we presented a novel approach to help developers create consistent and concise privacy captions from Android applications' source code. We described how using static analysis, we can identify code segments that process sensitive information and by using NMT models, we can translate these code segments into privacy captions. We also trained an NMT model on a code captioning dataset and used the model to generate captions for the PRCS. We evaluated the quality of the captions and found promising results. We also discussed the current challenges that we will address in the future.


\label{sec:conclusion} 

\bibliographystyle{unsrt}  
\bibliography{main}

\end{document}